\documentclass{aa}  

\usepackage{graphicx}
\usepackage[font=small,labelfont={bf,small}]{caption}
\usepackage[varg]{txfonts}
\usepackage{natbib,twoopt}
\usepackage[breaklinks=true]{hyperref} 

\bibpunct{(}{)}{;}{a}{}{,}             
\makeatletter
  \newcommandtwoopt{\citeads}[3][][]{\href{http://adsabs.harvard.edu/abs/#3}%
    {\def\hyper@linkstart##1##2{}%
     \let\hyper@linkend\@empty\citealp[#1][#2]{#3}}}
  \newcommandtwoopt{\citepads}[3][][]{\href{http://adsabs.harvard.edu/abs/#3}%
    {\def\hyper@linkstart##1##2{}%
     \let\hyper@linkend\@empty\citep[#1][#2]{#3}}}
  \newcommandtwoopt{\citetads}[3][][]{\href{http://adsabs.harvard.edu/abs/#3}%
    {\def\hyper@linkstart##1##2{}%
     \let\hyper@linkend\@empty\citet[#1][#2]{#3}}}
  \newcommandtwoopt{\citeyearads}[3][][]%
    {\href{http://adsabs.harvard.edu/abs/#3}
    {\def\hyper@linkstart##1##2{}%
     \let\hyper@linkend\@empty\citeyear[#1][#2]{#3}}}
\makeatother

\defcitealias{2007MNRAS.377.1187P}{P07}
\defcitealias{1973A&A....24..337S}{SS73}

\begin{document}

   \title{The ionizing and heating power of ultraluminous X-ray sources under the geometrical beaming model\thanks{Table~\ref{tab:average} is only available in electronic form
at the CDS via anonymous ftp to \url{cdsarc.u-strasbg.fr} (130.79.128.5)
or via \url{http://cdsweb.u-strasbg.fr/cgi-bin/qcat?J/A+A/}}}
    \titlerunning{The ionizing and heating power of ULXs under the geometrical beaming model}

   \author{
   K. Kovlakas\inst{1}\fnmsep\thanks{E-mail: konstantinos.kovlakas@unige.ch}
    \and T. Fragos\inst{1}
    \and D. Schaerer\inst{1,2}
    \and A. Mesinger\inst{3}
    }

   \authorrunning{Kovlakas et al.}
   
    \institute{
    Departement d’Astronomie, Université de Genève, Chemin Pegasi 51, CH-1290 Versoix, Switzerland
    \and
    CNRS, IRAP, 14 Avenue E. Berlin, 31400 Toulouse, France
    \and
    Scuola Normale Superiore, Piazza dei Cavalieri 7, I-56126 Pisa, Italy
    }
   
   \date{Received September 15, 1996; accepted March 16, 1997}


\abstract{
While there is now a consensus that X-ray binaries (XRBs) are the dominant X-ray sources in the early Universe and play a significant role during the epoch of heating of the intergalactic medium (IGM), recent studies report contradicting results regarding their contribution in the nebular emission of local Universe galaxies. Ultraluminous X-ray sources (ULXs), which dominate the X-ray budget of normal galaxies, may be important interstellar-medium (ISM) ionizing sources. However, their output in the extreme UV (EUV) and soft--X-ray part of the spectrum remains observationally unconstrained.
In this paper, we predict the ionizing and heating power from ULX populations under the geometrical beaming scenario, and three models describing the emission from super-critical accretion disks.
We find that our theoretical spectra for ULX populations cannot (can) explain the \ion{He}{II} (\ion{Ne}{V}) emission observed in some galaxies, with their contribution being less (more) important than the underlying stellar population. Stochastic fluctuations in the number of ULXs may allow for equal contributions in the \ion{He}{II} emission, in a fraction of galaxies.
We provide average spectra of ULX populations as an input to local, and early-Universe studies. We find that the soft--X-ray emission arising from super-critical accretion is significant for the heating of the IGM, and consistent with recent constraints from the 21-cm cosmic signal.
Based on the dependence on the adopted compact-object (CO) mass and accretion model, we encourage efforts in modeling ULX spectra via simulations, and their combination with detailed binary population synthesis models.
}


\maketitle


\section{Introduction}
\label{txt:introduction}

Low-metallicity galaxies have been found to exhibit high nebular \ion{He}{II} and \ion{Ne}{V} emission \citep[e.g.,][]{Garnett91,Izotov12}. The origin of this emission remains unknown, with multiple explanations in the recent literature: radiative shocks \citep[e.g.,][]{Plat19}, binary stars \citep[e.g.,][]{Gotberg17}, active galactic nuclei in dwarf galaxies \citep[e.g.,][]{Mezcua18}, XRBs \citep[e.g.,][]{Power13}, cluster winds \citep[e.g.,][]{Oskinova22}, etc. \citep[see discussion in][]{Olivier21}. Understanding ionizing sources is crucial for the study of the ISM \citep[e.g.,][]{Nanayakkara19}, and the heating of the IGM \citep[e.g.,][]{Mesinger13,Madau17}.

Observations show that XRB populations and their integrated luminosity are anticorrelated with the metallicity of the host galaxy \citep[e.g.,][]{Douna15}, also supported by theoretical work \citep[e.g.,][]{2013ApJ...764...41F,Fragos13}. This has motivated the study of the contribution of XRBs in the ionization of the ISM in low-metallicity galaxies, with contradicting results \citep[e.g.,][]{Schaerer19,Saxena20,Senchyna20,Umeda22}. In the case of the extremely low-metallicity galaxy I\,Zw\,18, exhibiting strong \ion{He}{II} emission, the X-ray luminosity is dominated by an ULX \citep[see review of][]{Kaaret17}, presenting an interesting case for the study of individual XRBs as ionizing sources. However, recent observational work is inconclusive: the source position and spectrum is not consistent with the \ion{He}{II} emission \citep[e.g.][]{Kehrig21}, unless a geometrical beaming effect is considered \citep[e.g.,][]{Rickards21}.

One of the most important difficulties in understanding the contribution of X-ray sources, which are typically observed at energy bands $0.2-10\,\rm keV$, is that their spectra are poorly constrained at the EUV regime (\ion{He}{II}: ${\sim}54\,\rm eV$, \ion{Ne}{V}: ${\sim}97.1\,\rm eV$; see \citealt{Lehmer22}). However, ULXs are strong candidates for four reasons: (i) ULXs exhibit softer spectra than typical XRBs and can potentially produce high rate of He-ionizing photons \citep[e.g.,][]{Simmonds21}, (ii) they have been shown to dominate the X-ray emission from normal galaxies due to the shallow slope of the high-mass XRB (HMXB) luminosity function \citep[XLF; e.g.,][]{Lehmer19}, (iii) they are abundant in low-metallicity environments \citep[e.g.,][]{Mapelli10}, and (iv) there is evidence of geometrical beaming \citep[e.g.,][]{2009MNRAS.393L..41K}, which indicates a larger underlying population of ULXs, and hence a stronger contribution in the ionization of the ISM than what is expected by the observed population \citep[e.g.,][]{Rickards21}.

An interesting case of a powerful ionizing source in the Galaxy is SS\,433 \citep{1977ApJS...33..459S}. It's high optical and UV emission \citep[e.g.,][]{2019A&A...624A.127W}, is believed to be powered by a super-critical accretion disk, despite its low observed X-ray luminosity \citep[e.g][]{2021MNRAS.506.1045M}. This can be explained through models of super-critical disks, which predict that close to the CO, the disk becomes geometrically thick, as strong optically-thick winds flow from its surface \citep[e.g.,][henceforth SS73]{1973A&A....24..337S}. The highly anisotropic radiation pattern results in face-on observers detecting collimated emission from the inner regions of the disk (beaming), and edge-on observers seeing \lq{}softer\rq{} emission escaping from the wind's photosphere extending to large radii with respect to the inner disk radius \citep[e.g.,][henceforth P07]{2007MNRAS.377.1187P}. Therefore, the common interpretation is that SS\,433 belongs to the general population of ULXs with observed luminosities $L_{\rm obs}{\gtrsim}10^{39}\,\rm erg\,s^{-1}$, but being viewed at a high inclination, its strong X-ray emission is invisible to us \citep[e.g.,][]{2006MNRAS.370..399B}. Consequently, a unified picture arises where the main difference between ULXs and soft ULXs is the viewing angle \citep[e.g.,][]{2013MNRAS.435.1758S,2017MNRAS.468.2865P}.

In this paper, we combine the properties of observed ULX populations and theoretical models describing their intrinsic spectra, and constrain the ionization power of ULXs under the super-Eddington accretion and beaming models.


\section{Methodology}

The contribution of ULX populations in the ionization of the ISM and the heating of the IGM is hampered by three observational challenges. Firstly, the hard EUV/soft--X-ray part of the spectrum cannot be detected directly due to absorption. Secondly, under the geometrical-beaming scenario, extragalactic edge-on ULXs are too faint to be detected by X-ray telescopes, and therefore, we may underestimate the ULX content of galaxies. Thirdly, the distributions of the physical properties of ULXs (e.g., accretor/donor masses, structure of the accretion disk, etc.) remain unknown. 
To overcome these challenges in our analysis, we adopt models for super-critically accreting sources that provide the luminosity, geometrical beaming and spectral energy distribution (SED), given the CO mass and mass-transfer rate of the systems. Below, we show that by considering different CO masses, the beaming factor and the spectrum depend only on the luminosity, and therefore, the underlying ULX population and its integrated spectra can be anchored on empirical constraints of the HMXB XLF.

\subsection{Luminosity distribution and beaming in ULXs}

Super-critical disks are, by definition, encountered in systems with luminosities exceeding the Eddington limit:
\begin{equation}
    L_{\rm Edd} = 1.26\times10^{38} m \ \rm erg\,s^{-1},
    \label{eq:eddington}
\end{equation}
where $m{=}M/M_\odot$ is the mass of the accreting object in solar units \citep[e.g.,][]{1980ApJ...242..772A}. Based on the typical range of BH masses in XRBs (${\lesssim}15\,M_\odot$; \citealt{Remillard06}) sources with luminosities exceeding $L_{\rm lim}{=}2{\times}10^{39}\,\rm erg\,s^{-1}$ are likely to be super-Eddington. Although super-critically accreting COs with near-solar masses may exhibit lower luminosities than the aforementioned limit, when we anchor our results on XLFs we ignore the part below $L_{\rm lim}$ to avoid mixing under- and super-Eddington sources.
In \citetalias{1973A&A....24..337S} it is shown that when the mass-transfer rate $\dot{M}$ exceeds the rate corresponding to the Eddington limit, $\dot{M}_{\rm Edd}$, the bolometric luminosity is
\begin{equation}
    L_{\rm bol} = L_{\rm Edd}\left(1 + \ln \dot{m}\right),
    \label{eq:SS73}
\end{equation}
where $\dot{m}{=}\dot{M}/\dot{M}_{\rm Edd}$. This is the result of a fraction of accretion power being spent for driving strong outflows, which keep the accretion on the CO Eddington-limited.
Even at extreme mass-transfer rates (e.g., $\dot{m}{\sim}1000$), the bolometric luminosity cannot reach $10^{41}\,\rm erg\,s^{-1}$ without invoking COs of $M{>}100\,M_\odot$. In the beaming scenario of \citet{2001ApJ...552L.109K}, the most extreme ULXs are explained by the fact that their observed X-ray luminosities ($L_{\rm obs}$) are the isotropic-equivalent of moderately super-Eddington, beamed sources viewed face-on:
\begin{equation}
    L_{\rm obs} = b^{-1} L_{\rm bol},
    \label{eq:beaming}
\end{equation}
where $b$ is the beaming factor. Furthermore, \citet{2009MNRAS.393L..41K} hinted at a dependence of the beaming factor on the mass-transfer rate,
\begin{equation}
    b = \begin{cases}
            1                                   & \dot{m} \leq 8.5 \\
            \left({8.5} / {\dot{m}}\right)^2    &  \dot{m} > 8.5
        \end{cases},
    \label{eq:beamingfactor}
\end{equation}
which explains the observed anticorrelation between the luminosity and temperature of the soft X-ray emission in ULXs \citep[e.g.,][]{2009MNRAS.398.1450K,2016ApJ...831..117F}.

We use the $L_{\rm obs}{>}2\times10^{39}\rm\,erg\,s^{-1}$ part of the HMXB XLF from \citet{Lehmer19} to model the distribution of ULX luminosities:
\begin{equation}
    \frac{dN}{dL_{38}} = {\rm SFR} \times K_{\rm HMXB} \times
        \begin{cases}
            L_{38}^{-\gamma}        & 20{<}L_{38}{<}L_c \\
            0                       & \mbox{elsewhere}
        \end{cases},
        \label{eq:lehmer}
\end{equation}
where $L_{38}{=}L_{\rm obs}/10^{38}\,\rm erg\,s^{-1}$, SFR is the star-formation rate of the parent stellar population, and the normalization $K_{\rm HMXB}$, power-law index $\gamma$ and cut-off luminosity $L_c$ are fitted parameters.
Equations \eqref{eq:eddington}--\eqref{eq:lehmer} interconnect the observable $L_{\rm obs}$ with the mass-transfer rate $\dot{m}$ via the XLF, assuming a distribution of masses $m$.
However, ULXs are rare sources \citep[typically one per galaxy; e.g.,][]{Kovlakas20}, and their distances are prohibiting in constraining their parameters, such as the mass of the accretor. For this reason, we will consider three different accretor masses: $1.4$ (corresponding to neutron stars\footnote{the NS mass is used as a lower limit; we do not account for presence of magnetic field/alternative accretion modes \citep[e.g.,][]{2021MNRAS.504..701B}.}), $8$ and $20\,M_\odot$.
Using Eqs. \eqref{eq:SS73}-\eqref{eq:beamingfactor} and the fact that by definition at sub-Eddington luminosities $L_{\rm bol}{=}\dot{m} L_{\rm Edd}$, we infer $\dot{m}$ from the ratio $f{=}L_{\rm obs}/L_{\rm Edd}$:
\begin{equation}
    \dot{m} = \begin{cases}
                f               & f \leq 1  \\
                {\rm e}^{f-1}         & 1 < f \leq{3.14} \\
                8.5\left(\frac{f}{3.14}\right)^{\frac{4}{9}}  & f > 3.14
            \end{cases}
    \label{eq:inverse}
    ,
\end{equation}
where the case for $f{>}1{+}\ln{8.5}{\simeq}3.14$ is an approximation of the inverse of $f{=}\left({\dot{m}}/{8.5}\right)^2\left(1{+}\ln\dot{m}\right)$ with accuracy ${<}2\%$ for $f{<800}$ (a sufficiently high value corresponding to $10^{41}\,\rm erg\,s^{-1}$ for $m{=}1$).

\subsection{The black-body component of the ULX spectrum}

The spectral properties of super-Eddington sources are highly dependent on the geometry of the disk, and the interplay between the outflowing gas and the radiation. Specifically, the wind's photosphere is expected to produce a soft component in the spectrum of ULXs \citep[e.g.,][]{2003MNRAS.345..657K}.

According to the \citetalias{1973A&A....24..337S} model, the local mass-transfer rate increases as we go towards the inner regions of the accretion disk. However, at super-Eddington rates, there is a radius, the spherization radius ($R_{\rm sp}$), at which the local Eddington limit is reached, initiating outflows. As a result, the mass-transfer rate decreases as we approach further the CO. At the $R_{\rm sp}$ the accretion disk is optically and geometrically thick, forming a nearly-spherical black-body source. The $R_{\rm sp}$ is computed by equalizing the accretion and Eddington luminosities \citep[cf.,][]{2002apa..book.....F}:
\begin{equation}
    R_{\rm sp} = \frac{G M \dot{M}}{\eta \dot{M}_{\rm Edd} c^2} = \frac{G M_\odot}{\eta c^2} m \dot{m} = 1.5\times 10^{5} m \dot{m} {\rm\,cm},
    \label{eq:Rsp}
\end{equation}
where $\eta$ is the accretion efficiency, for which we adopt the value of 0.1. 
Using the Stefan-Boltzmann law we calculate the effective temperature at the $R_{\rm sp}$:
\begin{equation}
    T_{\rm eff} = \left( \frac{L_{\rm Edd}}{4{\rm \pi} R_{\rm sp}^2 \sigma}  \right)^{\frac{1}{4}}
    \label{eq:Tsp}
    .
\end{equation}
However, this model might underestimate the radius of the black body. Theoretical work considering various physical processes in super-critical disks (for example wind and advection; e.g, \citealt{1999AstL...25..508L}, \citetalias{2007MNRAS.377.1187P}) have provided more realistic prescriptions for the temperature and radius of the photosphere from which a soft black-body component originates in ULXs (see review in \citealt{Fabrika21}). In \citetalias{2007MNRAS.377.1187P} the radius of the photosphere is estimated as:
\begin{equation}
    R_{\rm ph} = 3 \frac{e_w}{\zeta\beta} \dot{m}^{\frac{3}{2}} R_{\rm in},
    \label{eq:Rph}
\end{equation}
where $R_{\rm in}$ is the inner disk radius, and $e_w$, $\zeta$, and $\beta$ are model parameters (see below). We adopt an inner radius equal to three Schwarzschild radii for BHs (assuming non-spinning BHs):
\begin{equation}
    R_{\rm in} = \frac{6 G M_\odot}{c^2} m,
\end{equation}
and $11\rm\,km$ for NSs \citep{2016ApJ...820...28O}. The $T_{\rm eff}$ is
\begin{equation}
    T_{\rm ph} = 9.28\times 10^6 {\rm\,K} 
                 \left(\frac{\zeta \beta}{\epsilon_w}\right)^{\frac{1}{2}}
                 m^{-\frac{1}{4}} \dot{m}^{-\frac{3}{4}},
    \label{eq:Tph}
\end{equation}
where $\zeta=\sqrt{\xi^2-1}$ is a function of the ratio of the perpendicular velocity of the ejected gas over the orbital velocity ($\xi{=}u_z/u_K$), and $\beta$ is the ratio of the mean wind radial velocity over the orbital velocity at the $R_{\rm sp}$. Due to energy constraints (cf. \citetalias{2007MNRAS.377.1187P}), 
$\beta{\approx}1{\approx}\zeta$,
and therefore we vary only $e_w$, which is the fraction of the accretion energy powering the outflow. We adopt two values for $e_w$, 0.3 and 0.7 to study the effect of this parameter.

\subsection{The disk component of the ULX spectrum}

For all models, the soft black-body spectrum and its normalization is computed using Eqs. \eqref{eq:Rsp}, \eqref{eq:Tsp}, \eqref{eq:Rph}, and \eqref{eq:Tph}. The remaining emission is modeled by a multi-color disk black-body, using the \texttt{xspec} model \texttt{diskbb} from \texttt{sherpa} \citep{2007ASPC..376..543D}, which is parametrized by the inner temperature of the accretion disk, computed using the formula from \citetalias{2007MNRAS.377.1187P}:
\begin{equation}
    T_{\rm in} = 1.6 \times m^{-\frac{1}{4}}\left(1 - 0.2\dot{m}^{-\frac{1}{3}}\right) \ \text{keV}.
    \label{eq:Tin}
\end{equation}


\section{Results}

\subsection{Model predictions for the ionizing photons from ULXs}
\label{txt:resultssingle}

\begin{figure}
    \centering
    \includegraphics[width=0.97\columnwidth]{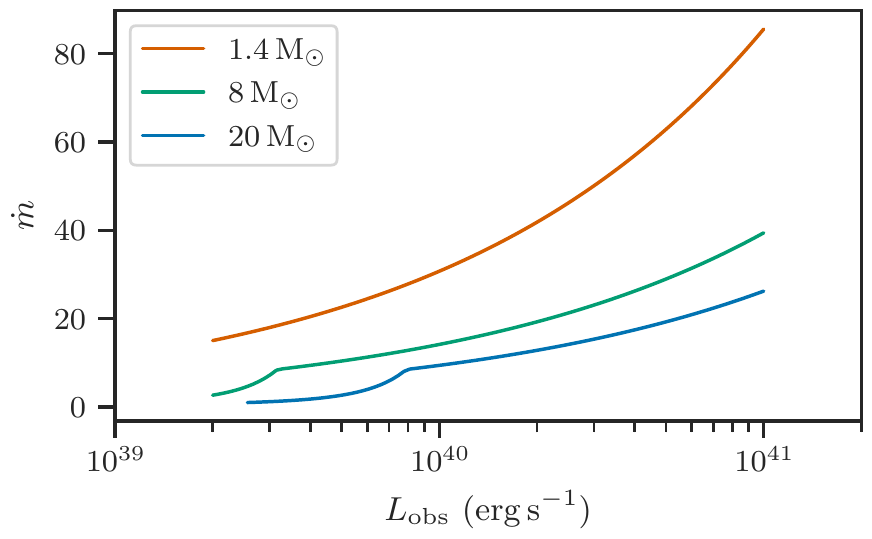}
    \includegraphics[width=0.97\columnwidth]{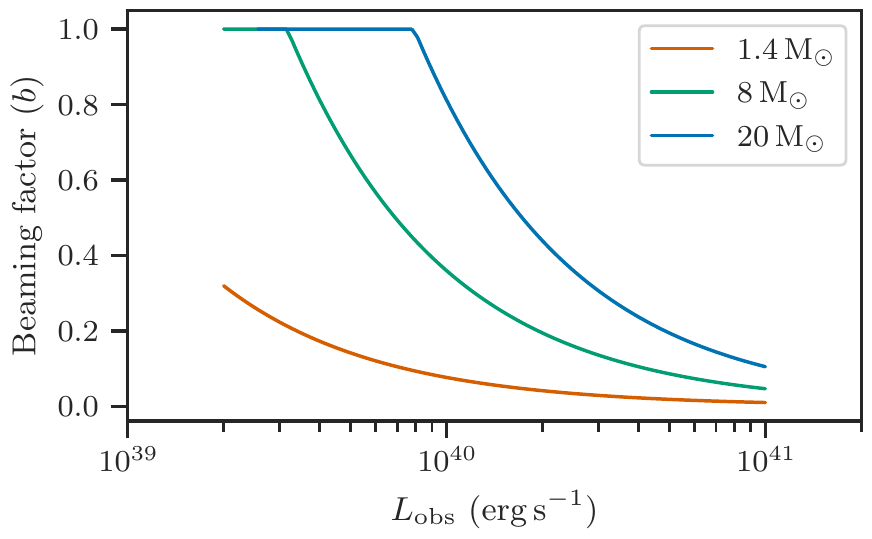}
    
    \includegraphics[width=0.97\columnwidth]{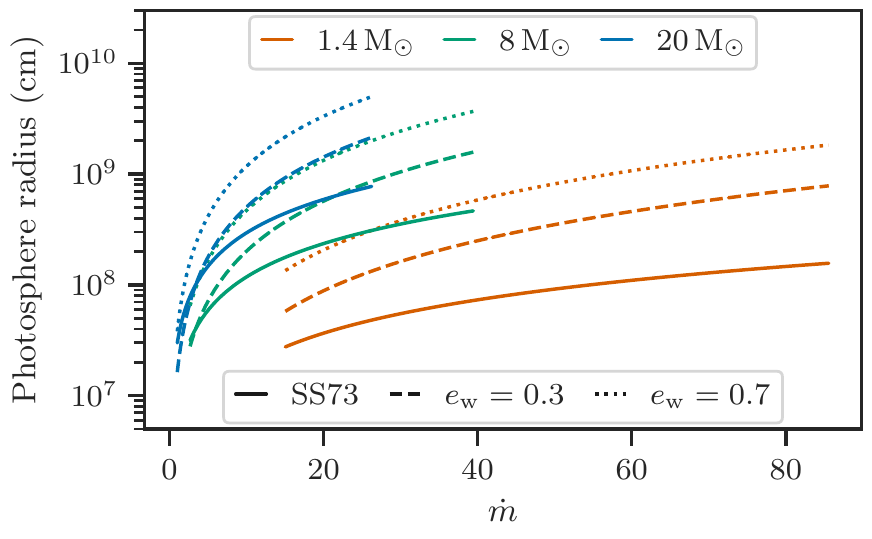}
    \includegraphics[width=0.97\columnwidth]{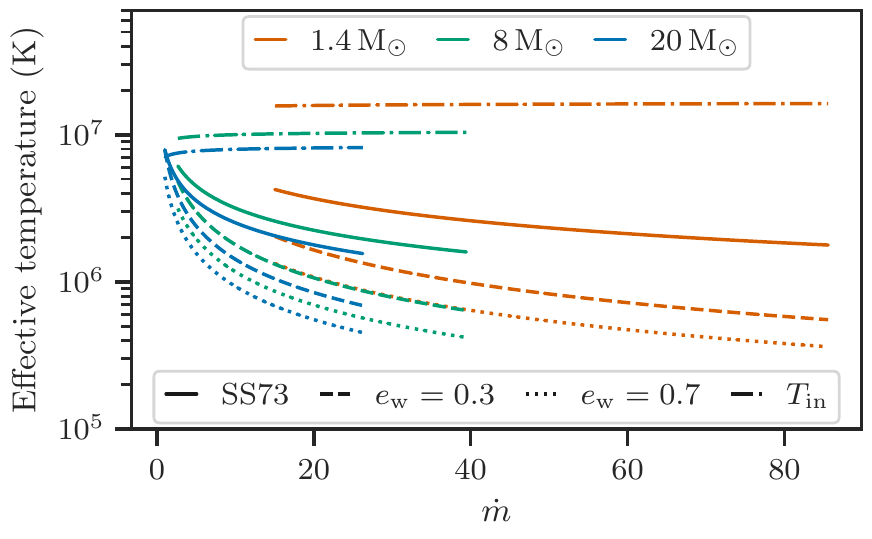}
    \caption{
    Solutions of Eqs. \eqref{eq:eddington}-\eqref{eq:Tin} for different CO masses (see legend) with $L_{\rm obs}$ in the range $2{-}100\times10^{39}\rm\,erg\,s^{-1}$. From top to bottom:
    \textbf{(i)}
    The mass-transfer rate ($\dot{m}$) of a ULX as a function of the isotropic-equivalent luminosity when observed face-on ($L_{\rm obs}$).
    \textbf{(ii)}
    The beaming factor ($b$) as a function of $L_{\rm obs}$; the flattening occurs at $\dot{m}{=}8.5$.
    \textbf{(iii)}
    The radius of the region producing the soft component as a function of $\dot{m}$, for different models (line styles; see lower legend).
    \textbf{(iv)}
    The effective temperature of the black-body region as a function of the $\dot{m}$; the dotted-dashed line shows the inner temperature of the disk.
    }
    \label{fig:4plots}
\end{figure}

For a grid of observed luminosities ($2{\times}10^{39}$--$10^{41}\,\rm erg\,s^{-1}$), and for three different CO masses (1.4, 8, and $20\,M_\odot$), using Eqs. \eqref{eq:eddington}-\eqref{eq:Tin}, we compute the $\dot{m}$ and $b$, as well as the photosphere radius and temperature for three cases: the \citetalias{1973A&A....24..337S} model, and the \citetalias{2007MNRAS.377.1187P} model with two values for the parameter $e_{\rm w}$ (see Fig.~\ref{fig:4plots}). As expected, the $T_{\rm ph}$ from the \citetalias{2007MNRAS.377.1187P} model is lower than the $T_{\rm eff}$ from \citetalias{1973A&A....24..337S} because of the larger photosphere radius compared to the $R_{\rm sp}$. We find that the total energy of the soft component from \citetalias{2007MNRAS.377.1187P} is ${\sim}L_{\rm Edd}/3$.

\begin{figure}
    \centering
    \includegraphics[width=\columnwidth]{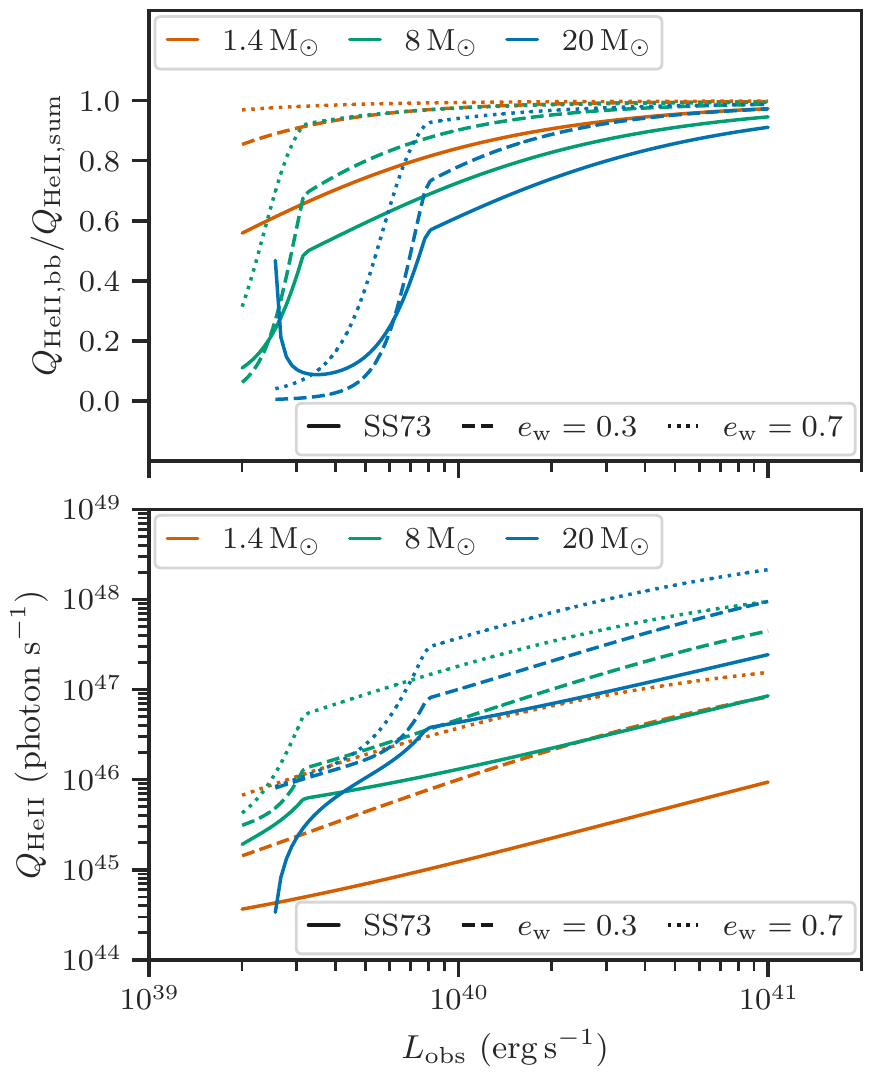}
    \caption{\ion{He}{II} ionization from ULXs with different CO masses (colors; see top legends) and spectrum models (line styles; see bottom legends), as a function of the face-on luminosity of the source.
        \textit{Top}:
            The fraction of the photons emerging from the black-body (bb) component with respect to the total (sum of black body and multi-color disk). 
        \textit{Bottom}:
            The rate of \ion{He}{II}-ionizing photons from the black-body and disk component, in the range $54.4{-}300\,\mathrm{eV}$.
    }
    \label{fig:Q_bb_disk}
\end{figure}

For three different ionization potentials, $E_{\rm ion}$ (13.6\,eV for \ion{H}{I}, 54.4\,eV for \ion{He}{II} and 96.6\,eV  for \ion{Ne}{V}), we compute the rate of ionizing photons 
$
    Q_{\rm ion}{=}\int_{E_{\rm ion}}^{E_{\rm max}} N_{\rm E}(E)\,dE,
$
where $N_E$ is the number of emitted photons with energy between $E$ and $E{+}dE$, and $E_{\rm max}{=}300\,\rm eV$. In Fig.~\ref{fig:Q_bb_disk} we show the number of \ion{He}{II}-ionizing photons as a function of the observed luminosity, and the relative contribution of the black-body component. The number of \ion{H}{I} and \ion{Ne}{V}-ionizing photons are higher by $0{-}5\%$, and lower by $5{-}20\%$, respectively, depending on the observed luminosity and the model.

\subsection{Comparison between ULX and stellar populations}

\begin{figure*}
    \centering
    \includegraphics[width=\columnwidth]{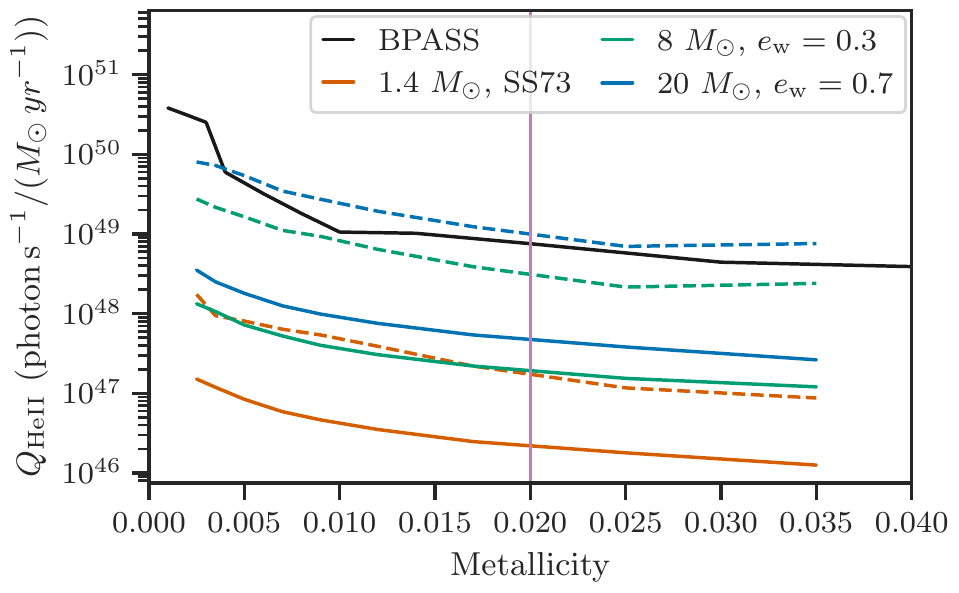}
    \includegraphics[width=\columnwidth]{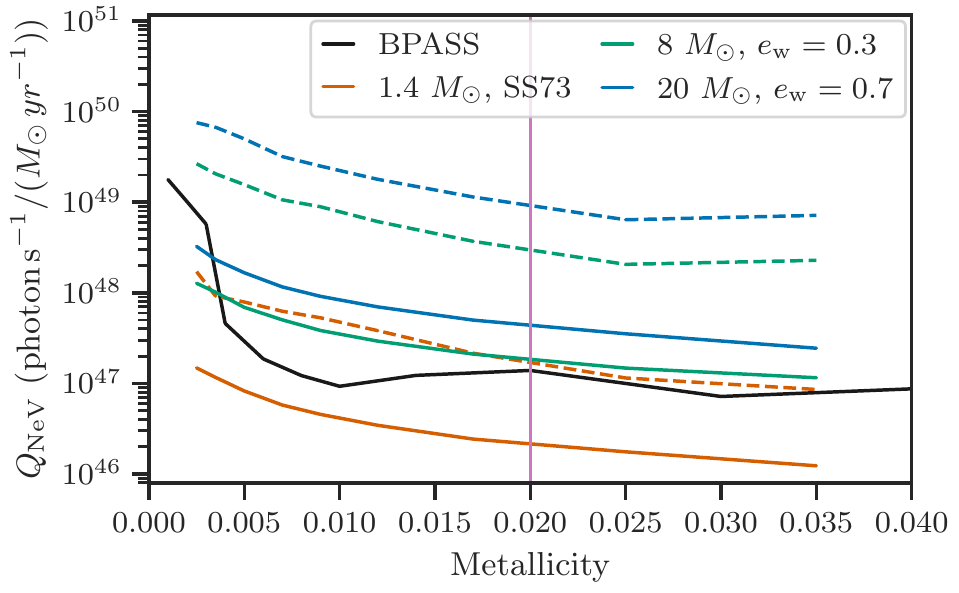}
    \caption{The rate of \ion{He}{II} (left) and \ion{Ne}{V} (right) ionizing photons from ULX populations in galaxies as a function of the gas-phase metallicity (solar value indicated with vertical magenta lines). The black lines show the prediction from stellar populations using \textit{BPASS} models. The colored solid lines depict the mean value from ULXs assuming three different combinations of CO masses and spectrum models (see legend), covering the full range of our (nine) estimates. The dashed lines show the upper 99\% limit to give a sense of the scatter due to the stochastic nature of ULXs. 
    \label{fig:BPASS_Q}
    }
\end{figure*}

We compute the rate of ionizing photons from underlying stellar populations using \textit{BPASS} \citep{2017PASA...34...58E} spectra for different metallicities (see Fig.~\ref{fig:BPASS_Q}), assuming continuous star-formation for 100\,Myr. We adopt the \textit{BPASS} value for the solar metallicity, $Z_\odot{=}0.02$.

The ionizing power of a ULX population in a given galaxy depends on the number of ULXs and their individual spectra. Our results on the rate of ionizing photons from ULXs as a function of the observed X-ray luminosity (\S\ref{txt:resultssingle}), in combination with XLFs \citep[e.g.,][]{Lehmer19}, provide a handle on the ionizing power of ULX populations.

To get a representative galaxy sample, such as the one used in the recent ULX demographic study of \citet{Kovlakas20}, we select all star-forming galaxies within a distance of $40\rm\,Mpc$, with reliable SFR and metallicity estimates in the Heraklion Extragalactic Catalogue (\textit{HECATE}; \citealt{2021MNRAS.506.1896K}). 
Since the \textit{HECATE} uses infrared SFR indicators and metallicities based on optical emission-line ratios, which are biased in the case of passive galaxies or in the presence of active galactic nuclei, we only select the 1,061 objects without nuclear activity, and \textit{SDSS} colors $g{-}r{<}0.65\,\rm mag$ (with a cut in the uncertainties in the photometry: $e_g, e_r{<}0.1\rm\,mag$; Kyritsis et al., in preparation). We use the \lq{}homogenized SFR indicator\rq{} which combines five different SFR indicators in the \textit{HECATE}. The SFR in this sample spans from $3.9\times 10^{-2}$ to $9.9\,M_\odot\,\rm yr^{-1}$, which is consistent with the range of SFRs in the calibration of the homogenized SFR indicator in the \textit{HECATE} (see fig. 5 in \citealt{2021MNRAS.506.1896K}), but most importantly falls in the region where the indicator is linearly correlated with SED-based SFR estimates for galaxies with $g{-}r{<}0.65$ (see fig. 6 in \citealt{2021MNRAS.506.1896K}). The metallicity\footnote{We convert the \textit{HECATE} gas-phase metallicities, $12{+}\log_{10}\left(O/H\right)$ to $Z$ using $12{+}\log_{10}\left(O/H\right)_\odot{=}8.69$ \citep{2009ARA&A..47..481A} and $Z_\odot{=}0.02$ \citep{2017PASA...34...58E}.} in our sample is in the range $\left[0.0025, 0.032\right]$.

For each galaxy, we sample the XLF (see Eq.~\ref{eq:lehmer}) above $2{\times}10^{39}\,\rm erg\,s^{-1}$. Each XLF is scaled for the SFR of the galaxy, and for its metallicity using the fit from \citet{Douna15}. However, we note that both the normalization and the shape of the HMXB XLF might depend on the metallicity \citep{Lehmer21}.

The XLFs are also corrected for the geometrical beaming. Under the beaming scenario, not all ULXs are observed due to the non-favorable angles towards us. While this has no effect on the integrated observed X-ray luminosity (the number of the sources is decreased by a factor of $b$, but their observed luminosity increases by $b$ as well), the intrinsic population of ULXs is underestimated. For a given CO mass, the beaming factor is a function of the luminosity (cf. Fig.~\ref{fig:4plots}) and therefore, the \lq{}unbeamed\rq{} XLF is:
$
    \frac{dN_{\text{corr}}(L_{38})}{dL_{38}} = b^{-1}(L_{38}) \frac{dN(L_{38})}{dL_{38}}
$.

We sample each galaxy's XLF 1000 times to quantify the stochasticity characterizing the high-end of the XLF. 
For each galaxy, we estimate the expected number of ULXs by integrating the XLF above $L_{\rm lim}$, and use this value as the mean of the Poisson distribution (typically $0.1{-}10$ ULXs depending on the properties of the galaxies) from which we sample the 1000 numbers of ULXs corresponding to each iteration. Then, for each iteration, we sample from the XLF to get a list of luminosities.
Using the results from \S\ref{txt:resultssingle} we sum the $Q_{\rm ion}$ of the ULXs (based on their observed luminosity; see bottom panel of Fig.~\ref{fig:Q_bb_disk}) in each iteration and galaxy, for all ionization potentials, CO masses, and models. In order to show the trend with the metallicity, we bin the galaxies by metallicity with bin edges at metallicity values that correspond to \textit{BPASS} results, 0.002, 0.003, 0.004, 0.006, 0.008, 0.01, 0.014, 0.02, 0.03, and 0.04, and compute the average rate of ionizing photons, as well as the 99\% percentile.

In Fig.~\ref{fig:BPASS_Q} we show the rate of ionizing photons for only three combinations of CO masses and models (everything else falls in-between), and only for the \ion{He}{II} and \ion{Ne}{V} (the Q for the \ion{H}{I} is smaller than the stellar one by ${\sim}4\,\rm dex$). The solid lines correspond to the mean value in the aforementioned metallicity bins, while the dashed ones, to the 99\% percentile showing how the stochastic nature of ULXs may lead to \ion{He}{II} emitters. 
Only a small fraction of galaxies, and under a favorable scenario (massive CO, $e_w{=}0.7$), host ULX populations with \ion{He}{II}-ionizing power comparable to that from the stellar populations. Depending on the CO mass and the adopted model, the \ion{Ne}{V}-ionizing power of ULX populations can match, or even exceed by 2\,dex that of the stellar populations. The scatter of an order of magnitude is caused by the stochastic nature of ULXs and the shape of the XLF. Specifically, in low-metallicty galaxies, which in general are characterized by low SFRs, the ULX content is small (${\sim}0.1$ ULXs) and therefore they often do not host a ULX. On the other hand, actively star-forming galaxies, with higher metallicity, are expected to host ${\gtrsim}1$ ULXs, but due to the shallow slope of the XLF, their luminosities and ionizing power covers a wide range of values.

\subsection{Average spectrum of ULX populations}
\label{txt:averagespectrum}

\begin{figure*}
    \centering
    \includegraphics[width=\textwidth]{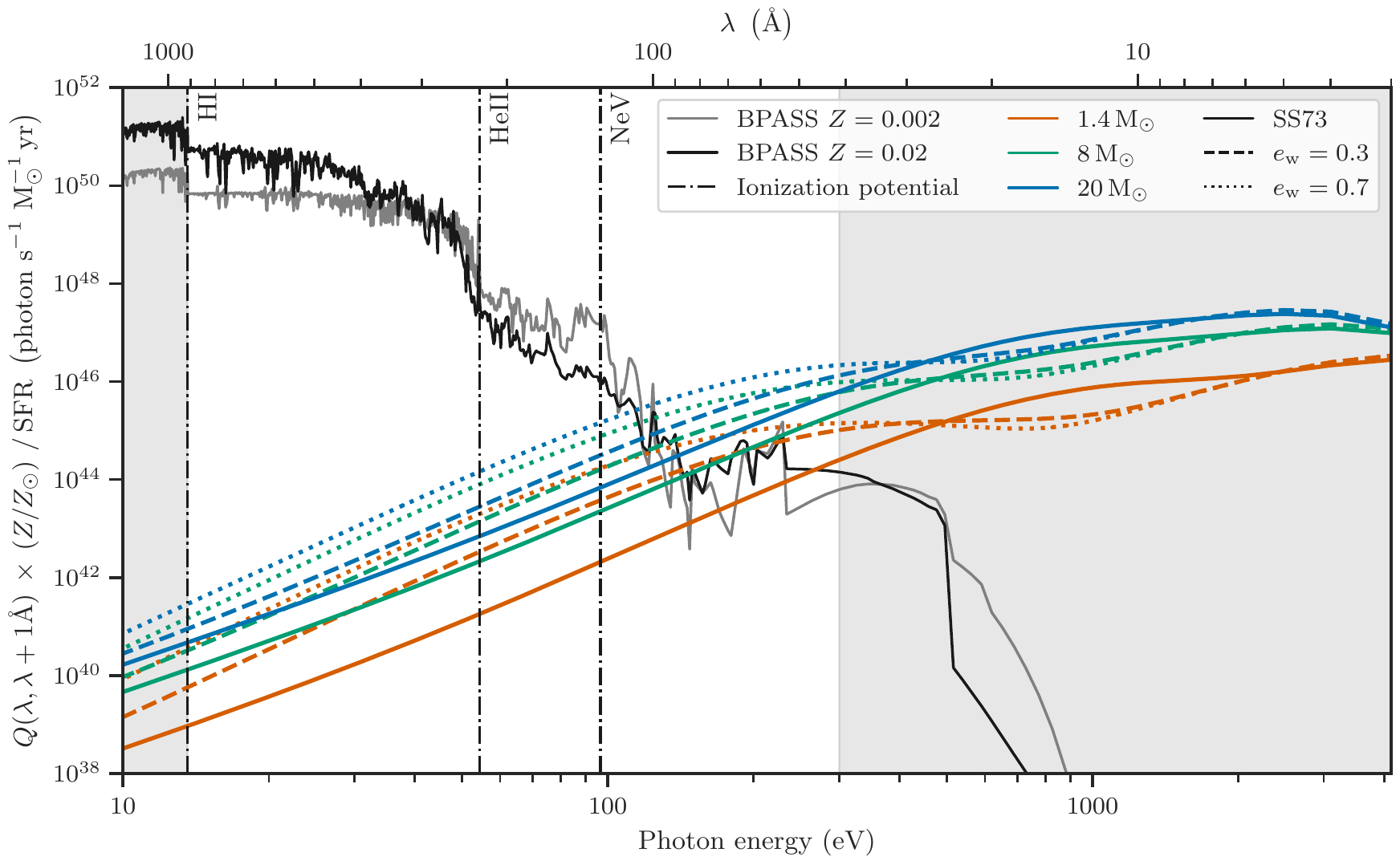}
    \caption{The average intrinsic spectrum from ULX populations assuming different CO masses (color) and spectrum models (line style), as well as stellar populations from \textit{BPASS} (black line for solar metallicity and gray for one-tenth of the solar metallicity). All spectra are normalized by the SFR, and the metallicity ($Z$); since we assume that the XLF scales with $Z_\odot/Z$, there is no variation in the ULX spectra, contrary to the stellar populations. The units in the $y$-axis are the same as the provided data (see Table~\ref{tab:average}). The ionization potentials of \ion{H}{I}, \ion{He}{II} and \ion{Ne}{V} are indicated with vertical lines. The shaded areas denote the regions in the spectrum that we do not consider as ionizing when calculating $Q_{\rm HI}$, $Q_{\rm HeII}$ and $Q_{\rm NeV}$ (see Figs.~\ref{fig:Q_bb_disk} and \ref{fig:BPASS_Q}).
    }
    \label{fig:avg_spectrum}
\end{figure*}

While in the previous section we focused on the ionizing power of stochastic ULX populations hosted in individual galaxies, here we construct a SFR-scaled and metallicity-dependent average spectrum of ULXs in star-forming galaxies that can be used as an input to IGM-heating and cosmic X-ray background studies \citep[e.g.,][]{2018ApJ...869..159U}.
To do so, we sum the XLF-weighted theoretical spectra for ULXs with $2{\times}10^{39}\,\rm erg\,s^{-1}{<}L_{\rm obs}{<}10^{41}\,\rm erg\,s^{-1}$. We also normalize the spectra for the metallicity since we assume that the XLF is linearly anticorrelated with the metallicity \citep[e.g.,][]{Douna15}. Fig.~\ref{fig:avg_spectrum} shows the spectra (see Table~\ref{tab:average} for the data) along with the stellar models from \textit{BPASS} for two different metallicity values. 

\begin{table*}
\centering
\small
\caption{Average ULX spectra scaled by the SFR and metallicity ($Z$), 
corresponding to all nine combinations of CO masses (1.4, 8, and $20\,M_\odot$), and black-body models (SS73, $e_w{=}0.3$, and $e_w{=}0.7$). The first and second columns are the photon energy ($E_{\rm ph}$) and wavelength ($\lambda$), respectively, while the rest are the decimal logarithm of the photon flux in the $(\lambda, \lambda+1\rm\AA$ bin for each combination of parameters. I.e., ${\rm SFR}\times(Z/Z_\odot)\times\sum_{i=1}^{10^4} 10^{y_i}{E_{\rm ph,i}}$, where $i$ is the row index and $y_i$ is a given model column, gives the total intrinsic luminosity (in eV) in the range $[1, 10^4]\rm \AA$ for a ULX population in an underlying stellar population of a given SFR and metallicity.
The full table with 100,000 entries is available in electronic form
at the CDS via anonymous ftp to \url{cdsarc.u-strasbg.fr} (130.79.128.5)
or via \url{http://cdsweb.u-strasbg.fr/cgi-bin/qcat?J/A+A/}.
}
\label{tab:average}
\begin{tabular}{|c|c|c|c|c|c|c|c|c|c|c|}
\hline
 \multicolumn{2}{|r|}{CO mass ($M_\odot$)} 
    & 1.4 & 1.4 & 1.4 
    & 8.0 & 8.0 & 8.0 
    & 20.0 & 20.0 & 20.0 \\
 \multicolumn{2}{|r|}{Model} 
    & SS73 & $e_w{=}0.3$ & $e_w{=}0.7$
    & SS73 & $e_w{=}0.3$ & $e_w{=}0.7$
    & SS73 & $e_w{=}0.3$ & $e_w{=}0.7$ \\\hline
$E_{\rm ph}$ & $\lambda$ & 
    \multicolumn{9}{|c|}{Decimal logarithm of photon flux in $\left(\lambda, \lambda+1\AA\right)$ bin, scaled by SFR and metallicity} \\
(eV) & (\AA) & 
    \multicolumn{9}{|c|}{$\rm (photon\,s^{-1}\,Myr^{-1}\,yr \times Z/Z_\odot)$}
\\\hline
0.123984 & 100000 & 32.11996 & 32.20639 & 32.21280 & 33.28700 & 33.39010 & 33.39158 & 33.84989 & 33.98321 & 33.98385 \\
0.123986 & 99999 & 32.11997 & 32.20640 & 32.21281 & 33.28701 & 33.39010 & 33.39159 & 33.84989 & 33.98321 & 33.98386 \\
0.123987 & 99998 & 32.11998 & 32.20640 & 32.21281 & 33.28702 & 33.39011 & 33.39160 & 33.84990 & 33.98322 & 33.98386 \\
0.123988 & 99997 & 32.12506 & 32.21149 & 32.21790 & 33.29210 & 33.39519 & 33.39668 & 33.85498 & 33.98830 & 33.98895 \\
0.123989 & 99995 & 32.11999 & 32.20642 & 32.21283 & 33.28703 & 33.39012 & 33.39161 & 33.84991 & 33.98323 & 33.98387 \\
\dotfill & \dotfill & \dotfill & \dotfill & \dotfill & \dotfill & \dotfill & \dotfill & \dotfill & \dotfill & \dotfill \\
\hline
\end{tabular}
\end{table*}

\subsection{Average effect on IGM heating}

\begin{figure}
    \centering
    \includegraphics[width=\columnwidth]{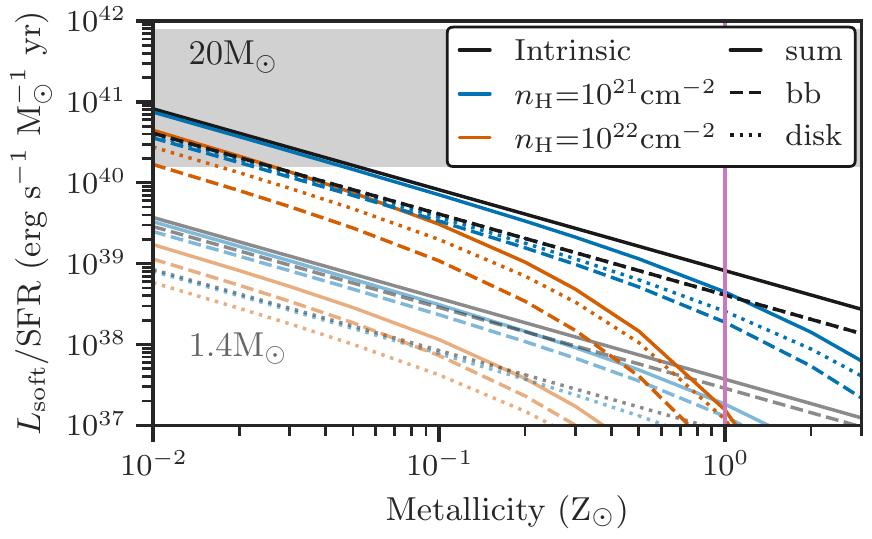}
    \caption{The average soft X-ray luminosity ($L_{\rm 0.3-1\,keV}$) of ULX populations as a function of the host galaxy's metallicity (solar value indicated with a vertical magenta line), assuming CO masses of $20\,M_\odot$ (top, intense lines) and $1.4\,M_\odot$ (bottom, pale lines) and spectrum described by the \citetalias{2007MNRAS.377.1187P} model with $e_{\rm w}{=}0.3$. Black lines correspond to the intrinsic (unabsorbed) X-ray luminosity, while blue and orange lines account for the absorption in the galaxy's ISM, with $n_{\rm H}{=}10^{21}$ and $10^{22}\,\rm cm^2$ respectively. Dashed, dotted and solid lines indicate the black-body component, disk component and their sum, respectively. (For $20\,M_\odot$, the black dashed and dotted lines coincide, indicating equal component contributions.) The gray band denotes the $L_{\rm X{<}2\,keV}$ 68\% confidence interval constrained from the cosmic 21-cm signal for galaxies at $z{\sim}8$ \citep{HERA22}.}
    \label{fig:softX}
\end{figure}

Using the average spectra in \S\ref{txt:averagespectrum}, we quantify the effect of ULX populations in IGM-heating at the epoch of cosmic heating. We focus on the $0.3{-}1\,\rm keV$ part of the intrinsic spectrum, since at lower energies the photons are absorbed by the ISM of the host galaxies, whereas higher energy photons are expected to penetrate the IGM and not contribute significantly in the heating. In addition, we study the effect of the metallicity-dependent photo-electric absorption of the ISM by applying the model \texttt{vphabs} from \texttt{xspec} for different values of hydrogen column density ($n_{\rm H}$).

For all CO masses, the soft X-ray luminosity as a function of metallicity depends weakly on the spectrum model (SS37, \citetalias{2007MNRAS.377.1187P}; within a factor two). In Fig.~\ref{fig:softX} we show the results for the two extreme CO masses (1.4 and $20\,M_\odot$) and for $e_w{=}0.3$. The normalization is roughly linear to the accretor mass, since the models predict harder spectra for lower CO masses (the black-body component is nearly Eddington-limited). 
While the exact distribution of CO masses is important to constrain the soft X-ray part of average ULX spectra, we find that the black-body and disk components have comparable contribution, with a factor of $1.1$--$9.3$ underestimation if we neglect the black-body component.


\section{Discussion}

Assuming that the spectra of ULXs are the sum of a black body (using the \citetalias{1973A&A....24..337S} and \citetalias{2007MNRAS.377.1187P} models) and a disk component, we quantify the ionizing power of individual ULXs for different CO masses. We find a wide range in $Q_{\rm HeII}{\sim} 10^{45-48}\rm\,photon\,s^{-1}$, which is considerably lower than the rates from He-emitters such as the I\,Zw\,18 galaxy, ${\sim}1.33{\times}10^{50}\rm\,photon\,s^{-1}$ \citep{Kehrig15}. The ULX in this galaxy has been investigated as the origin of the ionizing radiation \citep[e.g.,][]{Schaerer19}, however difficulties in constraining the EUV part of the spectrum of a ULX \citep[e.g.,][]{Simmonds21} might result in stark differences in its ionization power. Our theoretical ULX SEDs are not capable of producing strong \ion{He}{II} ionization ($Q_{\rm HeII}/L_{\rm X}{\leq}4{\times}10^{7} \rm photon\, erg^{-1}$) as invoked in previous studies 
(e.g., $2{\times}10^{10}\,\rm photon\,erg^{-1}$; \citealt{Schaerer19})

\begin{figure*}
    \centering
    \includegraphics[width=\textwidth]{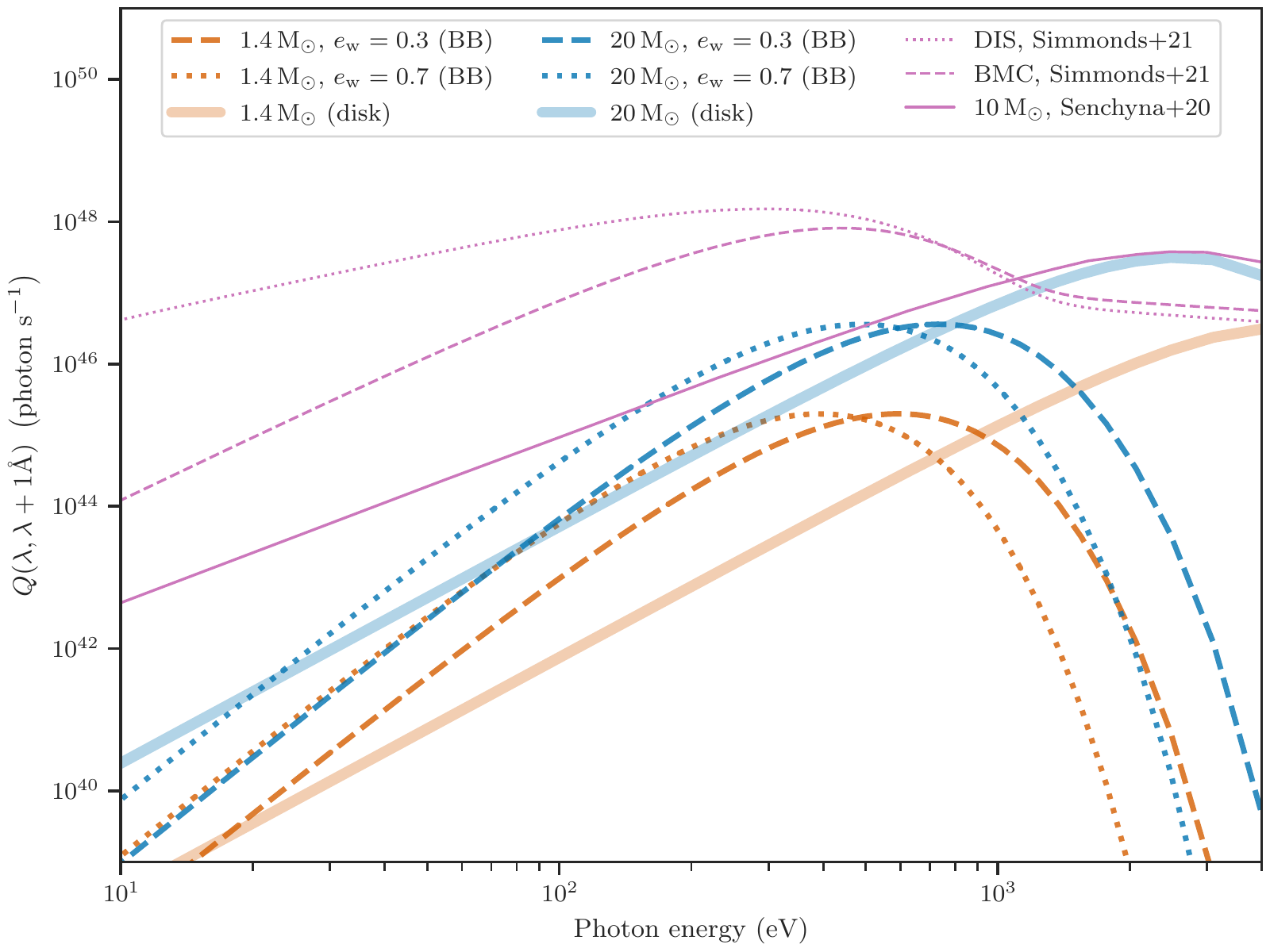}
    \caption{The disk (thick lines) and black-body (BB; thin lines) components of the intrinsic spectra of a ULX with $L_{\rm obs}{=}10^{40}\rm\,erg\,s^{-1}$ for two different CO masses, $1.4\,M_\odot$ (orange) and $20\,M_\odot$ (blue), and two different values of $e_{\rm w}$ for the \citetalias{2007MNRAS.377.1187P} model (line styles, see legend).
    For comparison, we overplot in purple the DIS (dotted line) and BMC (dashed line) models from \citet{Simmonds21}, as well as the model for a BH of mass $10\,M_\odot$ (continuous line) from \citet{Senchyna20}, scaled for the same source luminosity. 
    We note that in our models, the normalization of the intrinsic spectra depends on the CO mass: the smaller the CO mass is, the smaller the beaming factor is for the same observed, face-on luminosity.
    }
    \label{fig:compothers}
\end{figure*}

In Fig.~\ref{fig:compothers}, we compare the model spectra with previous studies, for a ULX with $L_{\rm obs}{=}10^{40}\,\rm erg\,s^{-1}$. Specifically, we show the spectrum of the disk and the black-body component of a ULX with observed (face-on) luminosity $10^{40}\rm\,erg\,s^{-1}$ for two different CO masses, and two different values of $e_{\rm w}$ in the \citetalias{2007MNRAS.377.1187P} model.
For the same observed, isotropic-equivalent luminosity, different CO masses correspond to different beaming factors. For this reason, in the case of the $1.4\,M_\odot$ accretor ($b{\simeq}0.076$), the bolometric luminosity of the source is lower than in the case of the $20\,M_\odot$ accretor ($b{\simeq}0.81$), leading to a lower normalization for the former.
While the shape of the black-body component is affected by the adopted value for $e_{\rm w}$, the disk component is the same for a given CO mass (the corresponding lines in Fig.~\ref{fig:compothers} are not repeated for different values of $e_{\rm w}$.)
We also show the models from \citet{Simmonds21} and \citet{Senchyna20}, scaled at the same luminosity, the normalization of which do not depend on the beaming factor, since geometrical collimation is not considered in these studies. We note that these models are overplotted to allow for qualitative comparisons against recent studies of the contribution of ULXs and XRBs in nebular emission.
Our individual ULX SEDs show that the photon flux at ${\sim}54.4\,\rm eV$ varies by one order of magnitude depending on the parameters of the models. However, they are weaker by many orders of magnitude compared to the literature models. We should stress again that no direct observations exist in this range, and in all cases, the ULX SEDs are extrapolated from theoretical or empirical models, calibrated to higher energies (${\gtrsim}300\,\rm eV$).

When considering the ionizing power of ULX populations, the geometrical beaming model results into two opposing effects. On one hand, the bolometric luminosity of the sources, and consequently their ionization power is lower than what is inferred from the observed luminosities. On the other hand, a fraction of the ULXs cannot be observed in the X-rays, but nevertheless ionizes the ISM. While the X-ray budget is preserved, the picture is more complicated in the EUV part of the spectrum: its shape and the beaming factor are correlated through their dependence on the $\dot{m}$. Consequently, the use of observed spectra and population synthesis techniques \citep[e.g.,][]{Fragos13} might underestimate the ionizing power of ULXs. Using the beaming-corrected HMXB XLF to anchor our analysis, we derive the underlying distribution of $\dot{m}$ and calculate the ionizing power of ULX populations in a realistic sample of galaxies. We find that in only a small fraction (${\sim}1\%$) of galaxies the ULXs compete the underlying stellar populations in \ion{He}{II} emission, in agreement with recent studies \citep[e.g.,][]{Senchyna20}, while they are significant in the case of \ion{Ne}{V} \citep[e.g.,][]{Simmonds21}. On the other hand, observational studies may also put constraints on the contribution of the hot gas component in star-forming galaxies, which is found to be comparable to the stellar component in ionizing the \ion{He}{II} in low-metallicity galaxies \citep[e.g.,][]{Lehmer22}.
From our modeling, the \ion{Ne}{V}-ionizing power of ULXs points at the possibility of constraints for the EUV emission from ULXs with the use of emission lines associated with high-ionization potentials. We encourage the study of high-ionization emission line galaxies, their emission processes, and in general systematic studies constraining the ionizing power and spectra of ULXs \citep[e.g.,][]{Izotov21}.

In addition, we compute the average spectrum of the ULX population in a galaxy normalized by its SFR and metallicity, for the different theoretical spectra, and adopted CO masses. These spectra can be used as an input for IGM-heating studies, and be compared to observational constraints from the cosmic 21-cm signal \citep[e.g.,][]{HERA22}. As an example, we calculate the heating power of ULX populations via their soft X-ray emission. We find that the contribution from the black-body component is comparable to the disk component, highlighting the importance of the former in quantifying the effect of ULXs in the early Universe. However, the normalization in this case depends strongly on the CO mass, and only weakly in the spectrum model among the ones investigated here.

Concluding, the dependence of our estimates on the shape of the spectra, the stochastic nature of ULXs, and the CO mass distribution, result in a 2\,dex scatter in the ionizing and heating power of ULX populations. Despite the success of analytical models in interpreting key properties of accreting sources, hydrodynamical simulations are necessary for investigating the radiative and mechanical feedback in super-critical accretion disks \citep[e.g.,][]{Sadowski15}.
Due to the computational cost of these simulations, only a handful of fiducial systems has been investigated. We encourage such efforts to continue, especially in the case of NS accretors where the presence of magnetic fields is important, with some sources showing lack of beaming \citep[e.g.,][]{Binder18}. Combining accurate spectral models with detailed binary population synthesis techniques \citep[e.g.,][]{2022arXiv220205892F} will provide stringent constraints on the ionizing power of ULXs, and their contribution in the heating of the early Universe.


\begin{acknowledgements}
We would like thank the anonymous referee for critical comments which improved the paper.
This work was supported by the Swiss National Science Foundation Professorship Grant (PP00P2\_176868; PI Fragos). KK acknowledges support from the Federal Commission for Scholarships for Foreign Students for the Swiss Government Excellence Scholarship (ESKAS No. 2021.0277). 
\end{acknowledgements}


\bibliographystyle{aa} 
\bibliography{main}

\begin{thebibliography}{58}
\expandafter\ifx\csname natexlab\endcsname\relax\def\natexlab#1{#1}\fi

\bibitem[{{Abdurashidova} {et~al.}(2022){Abdurashidova}, {Aguirre},
  {Alexander}, {Ali}, {Balfour}, {Barkana}, {Beardsley}, {Bernardi},
  {Billings}, {Bowman}, {Bradley}, {Bull}, {Burba}, {Carey}, {Carilli},
  {Cheng}, {DeBoer}, {Dexter}, {de Lera Acedo}, {Dillon}, {Ely}, {Ewall-Wice},
  {Fagnoni}, {Fialkov}, {Fritz}, {Furlanetto}, {Gale-Sides}, {Glendenning},
  {Gorthi}, {Greig}, {Grobbelaar}, {Halday}, {Hazelton}, {Heimersheim},
  {Hewitt}, {Hickish}, {Jacobs}, {Julius}, {Kern}, {Kerrigan}, {Kittiwisit},
  {Kohn}, {Kolopanis}, {Lanman}, {La Plante}, {Lekalake}, {Lewis}, {Liu}, {Ma},
  {MacMahon}, {Malan}, {Malgas}, {Maree}, {Martinot}, {Matsetela}, {Mesinger},
  {Mirocha}, {Molewa}, {Morales}, {Mosiane}, {Mu{\~n}oz}, {Murray}, {Neben},
  {Nikolic}, {Nunhokee}, {Parsons}, {Patra}, {Pieterse}, {Pober}, {Qin},
  {Razavi-Ghods}, {Reis}, {Ringuette}, {Robnett}, {Rosie}, {Santos}, {Sikder},
  {Sims}, {Smith}, {Syce}, {Thyagarajan}, {Williams}, \& {Zheng}}]{HERA22}
{Abdurashidova}, Z., {Aguirre}, J.~E., {Alexander}, P., {et~al.} 2022, \apj,
  924, 51

\bibitem[{{Abramowicz} {et~al.}(1980){Abramowicz}, {Calvani}, \&
  {Nobili}}]{1980ApJ...242..772A}
{Abramowicz}, M.~A., {Calvani}, M., \& {Nobili}, L. 1980, \apj, 242, 772

\bibitem[{{Asplund} {et~al.}(2009){Asplund}, {Grevesse}, {Sauval}, \&
  {Scott}}]{2009ARA&A..47..481A}
{Asplund}, M., {Grevesse}, N., {Sauval}, A.~J., \& {Scott}, P. 2009, \araa, 47,
  481

\bibitem[{{Begelman} {et~al.}(2006){Begelman}, {King}, \&
  {Pringle}}]{2006MNRAS.370..399B}
{Begelman}, M.~C., {King}, A.~R., \& {Pringle}, J.~E. 2006, \mnras, 370, 399

\bibitem[{{Binder} {et~al.}(2018){Binder}, {Levesque}, \&
  {Dorn-Wallenstein}}]{Binder18}
{Binder}, B., {Levesque}, E.~M., \& {Dorn-Wallenstein}, T. 2018, \apj, 863, 141

\bibitem[{{Brice} {et~al.}(2021){Brice}, {Zane}, {Turolla}, \&
  {Wu}}]{2021MNRAS.504..701B}
{Brice}, N., {Zane}, S., {Turolla}, R., \& {Wu}, K. 2021, \mnras, 504, 701

\bibitem[{{Doe} {et~al.}(2007){Doe}, {Nguyen}, {Stawarz}, {Refsdal},
  {Siemiginowska}, {Burke}, {Evans}, {Evans}, {McDowell}, {Houck}, \&
  {Nowak}}]{2007ASPC..376..543D}
{Doe}, S., {Nguyen}, D., {Stawarz}, C., {et~al.} 2007, in Astronomical Society
  of the Pacific Conference Series, Vol. 376, Astronomical Data Analysis
  Software and Systems XVI, ed. R.~A. {Shaw}, F.~{Hill}, \& D.~J. {Bell}, 543

\bibitem[{{Douna} {et~al.}(2015){Douna}, {Pellizza}, {Mirabel}, \&
  {Pedrosa}}]{Douna15}
{Douna}, V.~M., {Pellizza}, L.~J., {Mirabel}, I.~F., \& {Pedrosa}, S.~E. 2015,
  \aap, 579, A44

\bibitem[{{Eldridge} {et~al.}(2017){Eldridge}, {Stanway}, {Xiao}, {McClelland},
  {Taylor}, {Ng}, {Greis}, \& {Bray}}]{2017PASA...34...58E}
{Eldridge}, J.~J., {Stanway}, E.~R., {Xiao}, L., {et~al.} 2017, \pasa, 34, e058

\bibitem[{{Fabrika} {et~al.}(2021){Fabrika}, {Atapin}, {Vinokurov}, \&
  {Sholukhova}}]{Fabrika21}
{Fabrika}, S.~N., {Atapin}, K.~E., {Vinokurov}, A.~S., \& {Sholukhova}, O.~N.
  2021, Astrophysical Bulletin, 76, 6

\bibitem[{{Feng} {et~al.}(2016){Feng}, {Tao}, {Kaaret}, \&
  {Gris{\'e}}}]{2016ApJ...831..117F}
{Feng}, H., {Tao}, L., {Kaaret}, P., \& {Gris{\'e}}, F. 2016, \apj, 831, 117

\bibitem[{{Fragos} {et~al.}(2022){Fragos}, {Andrews}, {Bavera}, {Berry},
  {Coughlin}, {Dotter}, {Giri}, {Kalogera}, {Katsaggelos}, {Kovlakas},
  {Lalvani}, {Misra}, {Srivastava}, {Qin}, {Rocha}, {Roman-Garza}, {Serra},
  {Stahle}, {Sun}, {Teng}, {Trajcevski}, {Hai Tran}, {Xing}, {Zapartas}, \&
  {Zevin}}]{2022arXiv220205892F}
{Fragos}, T., {Andrews}, J.~J., {Bavera}, S.~S., {et~al.} 2022, arXiv e-prints,
  arXiv:2202.05892

\bibitem[{{Fragos} {et~al.}(2013{\natexlab{a}}){Fragos}, {Lehmer}, {Tremmel},
  {Tzanavaris}, {Basu-Zych}, {Belczynski}, {Hornschemeier}, {Jenkins},
  {Kalogera}, {Ptak}, \& {Zezas}}]{2013ApJ...764...41F}
{Fragos}, T., {Lehmer}, B., {Tremmel}, M., {et~al.} 2013{\natexlab{a}}, \apj,
  764, 41

\bibitem[{{Fragos} {et~al.}(2013{\natexlab{b}}){Fragos}, {Lehmer}, {Naoz},
  {Zezas}, \& {Basu-Zych}}]{Fragos13}
{Fragos}, T., {Lehmer}, B.~D., {Naoz}, S., {Zezas}, A., \& {Basu-Zych}, A.
  2013{\natexlab{b}}, \apjl, 776, L31

\bibitem[{{Frank} {et~al.}(2002){Frank}, {King}, \&
  {Raine}}]{2002apa..book.....F}
{Frank}, J., {King}, A., \& {Raine}, D.~J. 2002, {Accretion Power in
  Astrophysics: Third Edition}

\bibitem[{{Garnett} {et~al.}(1991){Garnett}, {Kennicutt}, {Chu}, \&
  {Skillman}}]{Garnett91}
{Garnett}, D.~R., {Kennicutt}, Robert~C., J., {Chu}, Y.-H., \& {Skillman},
  E.~D. 1991, \apj, 373, 458

\bibitem[{{G{\"o}tberg} {et~al.}(2017){G{\"o}tberg}, {de Mink}, \&
  {Groh}}]{Gotberg17}
{G{\"o}tberg}, Y., {de Mink}, S.~E., \& {Groh}, J.~H. 2017, \aap, 608, A11

\bibitem[{{Izotov} {et~al.}(2021){Izotov}, {Thuan}, \& {Guseva}}]{Izotov21}
{Izotov}, Y.~I., {Thuan}, T.~X., \& {Guseva}, N.~G. 2021, \mnras, 508, 2556

\bibitem[{{Izotov} {et~al.}(2012){Izotov}, {Thuan}, \& {Privon}}]{Izotov12}
{Izotov}, Y.~I., {Thuan}, T.~X., \& {Privon}, G. 2012, \mnras, 427, 1229

\bibitem[{{Kaaret} {et~al.}(2017){Kaaret}, {Feng}, \& {Roberts}}]{Kaaret17}
{Kaaret}, P., {Feng}, H., \& {Roberts}, T.~P. 2017, \araa, 55, 303

\bibitem[{{Kajava} \& {Poutanen}(2009)}]{2009MNRAS.398.1450K}
{Kajava}, J. J.~E. \& {Poutanen}, J. 2009, \mnras, 398, 1450

\bibitem[{{Kehrig} {et~al.}(2021){Kehrig}, {Guerrero}, {V{\'\i}lchez}, \&
  {Ramos-Larios}}]{Kehrig21}
{Kehrig}, C., {Guerrero}, M.~A., {V{\'\i}lchez}, J.~M., \& {Ramos-Larios}, G.
  2021, \apjl, 908, L54

\bibitem[{{Kehrig} {et~al.}(2015){Kehrig}, {V{\'\i}lchez}, {P{\'e}rez-Montero},
  {Iglesias-P{\'a}ramo}, {Brinchmann}, {Kunth}, {Durret}, \& {Bayo}}]{Kehrig15}
{Kehrig}, C., {V{\'\i}lchez}, J.~M., {P{\'e}rez-Montero}, E., {et~al.} 2015,
  \apjl, 801, L28

\bibitem[{{King}(2009)}]{2009MNRAS.393L..41K}
{King}, A.~R. 2009, \mnras, 393, L41

\bibitem[{{King} {et~al.}(2001){King}, {Davies}, {Ward}, {Fabbiano}, \&
  {Elvis}}]{2001ApJ...552L.109K}
{King}, A.~R., {Davies}, M.~B., {Ward}, M.~J., {Fabbiano}, G., \& {Elvis}, M.
  2001, \apjl, 552, L109

\bibitem[{{King} \& {Pounds}(2003)}]{2003MNRAS.345..657K}
{King}, A.~R. \& {Pounds}, K.~A. 2003, \mnras, 345, 657

\bibitem[{{Kovlakas} {et~al.}(2021){Kovlakas}, {Zezas}, {Andrews}, {Basu-Zych},
  {Fragos}, {Hornschemeier}, {Kouroumpatzakis}, {Lehmer}, \&
  {Ptak}}]{2021MNRAS.506.1896K}
{Kovlakas}, K., {Zezas}, A., {Andrews}, J.~J., {et~al.} 2021, \mnras, 506, 1896

\bibitem[{{Kovlakas} {et~al.}(2020){Kovlakas}, {Zezas}, {Andrews}, {Basu-Zych},
  {Fragos}, {Hornschemeier}, {Lehmer}, \& {Ptak}}]{Kovlakas20}
{Kovlakas}, K., {Zezas}, A., {Andrews}, J.~J., {et~al.} 2020, \mnras, 498, 4790

\bibitem[{{Lehmer} {et~al.}(2021){Lehmer}, {Eufrasio}, {Basu-Zych}, {Doore},
  {Fragos}, {Garofali}, {Kovlakas}, {Williams}, {Zezas}, \&
  {Santana-Silva}}]{Lehmer21}
{Lehmer}, B.~D., {Eufrasio}, R.~T., {Basu-Zych}, A., {et~al.} 2021, \apj, 907,
  17

\bibitem[{{Lehmer} {et~al.}(2022){Lehmer}, {Eufrasio}, {Basu-Zych}, {Garofali},
  {Gilbertson}, {Mesinger}, \& {Yukita}}]{Lehmer22}
{Lehmer}, B.~D., {Eufrasio}, R.~T., {Basu-Zych}, A., {et~al.} 2022, \apj, 930,
  135

\bibitem[{{Lehmer} {et~al.}(2019){Lehmer}, {Eufrasio}, {Tzanavaris},
  {Basu-Zych}, {Fragos}, {Prestwich}, {Yukita}, {Zezas}, {Hornschemeier}, \&
  {Ptak}}]{Lehmer19}
{Lehmer}, B.~D., {Eufrasio}, R.~T., {Tzanavaris}, P., {et~al.} 2019, \apjs,
  243, 3

\bibitem[{{Lipunova}(1999)}]{1999AstL...25..508L}
{Lipunova}, G.~V. 1999, Astronomy Letters, 25, 508

\bibitem[{{Madau} \& {Fragos}(2017)}]{Madau17}
{Madau}, P. \& {Fragos}, T. 2017, \apj, 840, 39

\bibitem[{{Mapelli} {et~al.}(2010){Mapelli}, {Ripamonti}, {Zampieri}, {Colpi},
  \& {Bressan}}]{Mapelli10}
{Mapelli}, M., {Ripamonti}, E., {Zampieri}, L., {Colpi}, M., \& {Bressan}, A.
  2010, \mnras, 408, 234

\bibitem[{{Mesinger} {et~al.}(2013){Mesinger}, {Ferrara}, \&
  {Spiegel}}]{Mesinger13}
{Mesinger}, A., {Ferrara}, A., \& {Spiegel}, D.~S. 2013, \mnras, 431, 621

\bibitem[{{Mezcua} {et~al.}(2018){Mezcua}, {Civano}, {Marchesi}, {Suh},
  {Fabbiano}, \& {Volonteri}}]{Mezcua18}
{Mezcua}, M., {Civano}, F., {Marchesi}, S., {et~al.} 2018, \mnras, 478, 2576

\bibitem[{{Middleton} {et~al.}(2021){Middleton}, {Walton}, {Alston}, {Dauser},
  {Eikenberry}, {Jiang}, {Fabian}, {Fuerst}, {Brightman}, {Marshall}, {Parker},
  {Pinto}, {Harrison}, {Bachetti}, {Altamirano}, {Bird}, {Perez},
  {Miller-Jones}, {Charles}, {Boggs}, {Christensen}, {Craig}, {Forster},
  {Grefenstette}, {Hailey}, {Madsen}, {Stern}, \&
  {Zhang}}]{2021MNRAS.506.1045M}
{Middleton}, M.~J., {Walton}, D.~J., {Alston}, W., {et~al.} 2021, \mnras, 506,
  1045

\bibitem[{{Nanayakkara} {et~al.}(2019){Nanayakkara}, {Brinchmann}, {Boogaard},
  {Bouwens}, {Cantalupo}, {Feltre}, {Kollatschny}, {Marino}, {Maseda},
  {Matthee}, {Paalvast}, {Richard}, \& {Verhamme}}]{Nanayakkara19}
{Nanayakkara}, T., {Brinchmann}, J., {Boogaard}, L., {et~al.} 2019, \aap, 624,
  A89

\bibitem[{{Olivier} {et~al.}(2021){Olivier}, {Berg}, {Chisholm}, {Erb},
  {Pogge}, \& {Skillman}}]{Olivier21}
{Olivier}, G.~M., {Berg}, D.~A., {Chisholm}, J., {et~al.} 2021, arXiv e-prints,
  arXiv:2109.06725

\bibitem[{{Oskinova} \& {Schaerer}(2022)}]{Oskinova22}
{Oskinova}, L.~M. \& {Schaerer}, D. 2022, \aap, 661, A67

\bibitem[{{{\"O}zel} {et~al.}(2016){{\"O}zel}, {Psaltis}, {G{\"u}ver}, {Baym},
  {Heinke}, \& {Guillot}}]{2016ApJ...820...28O}
{{\"O}zel}, F., {Psaltis}, D., {G{\"u}ver}, T., {et~al.} 2016, \apj, 820, 28

\bibitem[{{Pinto} {et~al.}(2017){Pinto}, {Alston}, {Soria}, {Middleton},
  {Walton}, {Sutton}, {Fabian}, {Earnshaw}, {Urquhart}, {Kara}, \&
  {Roberts}}]{2017MNRAS.468.2865P}
{Pinto}, C., {Alston}, W., {Soria}, R., {et~al.} 2017, \mnras, 468, 2865

\bibitem[{{Plat} {et~al.}(2019){Plat}, {Charlot}, {Bruzual}, {Feltre},
  {Vidal-Garc{\'\i}a}, {Morisset}, {Chevallard}, \& {Todt}}]{Plat19}
{Plat}, A., {Charlot}, S., {Bruzual}, G., {et~al.} 2019, \mnras, 490, 978

\bibitem[{{Poutanen} {et~al.}(2007){Poutanen}, {Lipunova}, {Fabrika},
  {Butkevich}, \& {Abolmasov}}]{2007MNRAS.377.1187P}
{Poutanen}, J., {Lipunova}, G., {Fabrika}, S., {Butkevich}, A.~G., \&
  {Abolmasov}, P. 2007, \mnras, 377, 1187

\bibitem[{{Power} {et~al.}(2013){Power}, {James}, {Combet}, \&
  {Wynn}}]{Power13}
{Power}, C., {James}, G., {Combet}, C., \& {Wynn}, G. 2013, \apj, 764, 76

\bibitem[{{Remillard} \& {McClintock}(2006)}]{Remillard06}
{Remillard}, R.~A. \& {McClintock}, J.~E. 2006, \araa, 44, 49

\bibitem[{{Rickards Vaught} {et~al.}(2021){Rickards Vaught}, {Sandstrom}, \&
  {Hunt}}]{Rickards21}
{Rickards Vaught}, R.~J., {Sandstrom}, K.~M., \& {Hunt}, L.~K. 2021, \apjl,
  911, L17

\bibitem[{{Saxena} {et~al.}(2020){Saxena}, {Pentericci}, {Schaerer},
  {Schneider}, {Amorin}, {Bongiorno}, {Calabr{\`o}}, {Castellano}, {Cimatti},
  {Cullen}, {Fontana}, {Fynbo}, {Hathi}, {McLeod}, {Talia}, \&
  {Zamorani}}]{Saxena20}
{Saxena}, A., {Pentericci}, L., {Schaerer}, D., {et~al.} 2020, \mnras, 496,
  3796

\bibitem[{{Schaerer} {et~al.}(2019){Schaerer}, {Fragos}, \&
  {Izotov}}]{Schaerer19}
{Schaerer}, D., {Fragos}, T., \& {Izotov}, Y.~I. 2019, \aap, 622, L10

\bibitem[{{Senchyna} {et~al.}(2020){Senchyna}, {Stark}, {Mirocha}, {Reines},
  {Charlot}, {Jones}, \& {Mulchaey}}]{Senchyna20}
{Senchyna}, P., {Stark}, D.~P., {Mirocha}, J., {et~al.} 2020, \mnras, 494, 941

\bibitem[{{Shakura} \& {Sunyaev}(1973)}]{1973A&A....24..337S}
{Shakura}, N.~I. \& {Sunyaev}, R.~A. 1973, \aap, 500, 33

\bibitem[{{Simmonds} {et~al.}(2021){Simmonds}, {Schaerer}, \&
  {Verhamme}}]{Simmonds21}
{Simmonds}, C., {Schaerer}, D., \& {Verhamme}, A. 2021, \aap, 656, A127

\bibitem[{{S{\k{a}}dowski} \& {Narayan}(2015)}]{Sadowski15}
{S{\k{a}}dowski}, A. \& {Narayan}, R. 2015, \mnras, 453, 3213

\bibitem[{{Stephenson} \& {Sanduleak}(1977)}]{1977ApJS...33..459S}
{Stephenson}, C.~B. \& {Sanduleak}, N. 1977, \apjs, 33, 459

\bibitem[{{Sutton} {et~al.}(2013){Sutton}, {Roberts}, \&
  {Middleton}}]{2013MNRAS.435.1758S}
{Sutton}, A.~D., {Roberts}, T.~P., \& {Middleton}, M.~J. 2013, \mnras, 435,
  1758

\bibitem[{{Umeda} {et~al.}(2022){Umeda}, {Ouchi}, {Nakajima}, {Isobe},
  {Aoyama}, {Harikane}, {Ono}, \& {Matsumoto}}]{Umeda22}
{Umeda}, H., {Ouchi}, M., {Nakajima}, K., {et~al.} 2022, \apj, 930, 37

\bibitem[{{Upton Sanderbeck} {et~al.}(2018){Upton Sanderbeck}, {McQuinn},
  {D'Aloisio}, \& {Werk}}]{2018ApJ...869..159U}
{Upton Sanderbeck}, P.~R., {McQuinn}, M., {D'Aloisio}, A., \& {Werk}, J.~K.
  2018, \apj, 869, 159

\bibitem[{{Waisberg} {et~al.}(2019){Waisberg}, {Dexter}, {Olivier-Petrucci},
  {Dubus}, \& {Perraut}}]{2019A&A...624A.127W}
{Waisberg}, I., {Dexter}, J., {Olivier-Petrucci}, P., {Dubus}, G., \&
  {Perraut}, K. 2019, \aap, 624, A127

\end{thebibliography}

\end{document}